\documentclass[fleqn,usenatbib]{mnras}
\usepackage{hyperref}
\hypersetup{
    colorlinks=true,
    linkcolor=blue,
    filecolor=magenta,
    urlcolor=cyan,
    pdfpagemode=FullScreen,
    }

\usepackage{newtxtext,newtxmath}
\usepackage[T1]{fontenc}
\usepackage{ae,aecompl}
\usepackage{graphicx}
\usepackage{amsmath}
\usepackage{lipsum}

\usepackage{multirow}
\usepackage{dcolumn}
\usepackage{orcidlink}

\def\lsim{~\rlap{$<$}{\lower 1.0ex\hbox{$\sim$}}}
\def\gsim{~\rlap{$>$}{\lower 1.0ex\hbox{$\sim$}}}

\title[Heavy seed survivors in dwarf galaxies]{Heavy black hole seed survivors in dwarf galaxies: a case study of Leo~I}
\author[M.~T.~Scoggins et al]{Matthew~T.~Scoggins\orcidlink{0000-0002-0748-9115},$^{1}$\thanks{E-mail: mts2188@columbia.edu}
\ Zolt{\'{a}}n~Haiman,$^{1,2,3}$
\ and Fabio~Pacucci$^{4,5}$
\\
$^{1}$Department of Astronomy, Columbia University, New York, NY, 10027\\
$^{2}$Department of Physics, Columbia University, New York, NY, 10027\\
$^{3}$Institute of Science and Technology Austria (ISTA), Am Campus 1, Klosterneuburg, Austria\\
$^{4}$Center for Astrophysics $\vert$ Harvard \& Smithsonian, 60 Garden St, Cambridge, MA 02138, USA\\
$^{5}$Black Hole Initiative, Harvard University, 20 Garden St, Cambridge, MA 02138, USA\\
}

\date{Accepted XXX. Received YYY; in original form ZZZ}

\pubyear{2025}

\begin{document}
\label{firstpage}
\pagerange{\pageref{firstpage}--\pageref{lastpage}}
\maketitle


\begin{abstract}
   The supermassive black holes (SMBHs) with mass $M_\bullet > 10^9 \, \rm M_\odot$ hosted by high-redshift galaxies have challenged our understanding of black hole formation and growth, as several pathways have emerged attempting to explain their existence. The "heavy-seed" pathway eases the problem with the progenitors of these SMBHs having birth masses up to ${\sim} 10^5~{\rm M_\odot}$. Here, we investigate the possibility that a local dwarf galaxy, Leo~I, harbors a heavy-seed descendant. Using Monte-Carlo merger trees to generate the merger histories of 1,000 dark matter halos similar to the Milky Way (MW; with a mass of ${\sim} 10^{12}~{\rm M_\odot}$ at redshift $z{=}0$). We search for Leo-like satellite halos among these merger trees, and investigate the probability that the progenitors of some of these satellites formed a heavy seed. We derive the likelihood of such "heavy seed survivors" (HSSs) across various formation and survival criteria as well as Leo-similarity criteria. We find that the virial temperature for the onset of atomic cooling and rapid gas infall that yields heavy seeds, $T_{\rm act}$, has the largest impact on the number of HSSs. We find HSSs in a fraction $0.7\%$, $18.1\%$, and $96.5\%$ of MW-like halos when $T_{\rm act}$ is set to $9,000$K, $7,000$K, and $5,000$K respectively. This suggests that Leo~I could be hosting a heavy seed and could provide an opportunity to disentangle heavy seeds from other SMBH formation mechanisms.
\end{abstract}

\section{Introduction}
\label{sec:intro}
The origin of supermassive black holes (SMBHs) heavier than $\sim10^9~{\rm M_\odot}$ powering quasars at redshift $z\geq 6$ remains poorly understood \citep{Fan_2001, Fan_2003, Mortlock_2011, Wu_2015, Banados_2018, Yang_2020, Wang_2021}. There are more than 200 detections of these SMBHs (for recent compilations, see \citealt{Fan_2022} and \citealt{Bosman_2022}). 

The James Webb Space Telescope has recently unveiled a new class of high-$z$ galaxies, named the Little Red Dots (LRDs; \citealt{Kocevski_2023,Harikane_2023, Matthee_2023, Kokorev_2024,Maiolino_2024,Baggen_2024,Guia_2024, Kocevski_2025}). They are compact high-redshift galaxies thought to host massive black holes, which appear to be overmassive with respect to the local standard relations, suggesting that early black holes got a head start compared to their host galaxies (\citealt{Pacucci_2023}, but see \citealt{Li_2025} for a counterargument to this claim). 

Understanding high-$z$ black hole formation and evolution requires a new understanding of the mechanisms for rapid and sustained black hole formation and growth. The solutions to this problem typically fall into two categories: 'light' and 'heavy' seeds \citep{Tanaka_2009, Volonteri_2010, Johnson_2016,Inayoshi_2020, Volonteri_2021,Regan_2024}. 

The collapse of gas in primordial dark matter (DM) halos is the precursor to the formation of astrophysical objects in the early universe. This collapse leads to fragmentation, which forms the first stars. For smaller 'minihalos' with virial temperature $T_{\rm vir} {<}10^4$K, this leads to the formation of Population III (Pop III) stars (each with mass of $M {\sim} 10{-}100 {\rm M_\odot}$; \citealt{Abel_2000, Bromm_2001, Abel_2002, Hirano_2014, Katz_2025}), which then go on to form 'light' seeds. In some (rare) circumstances, gas inside some minihalos can avoid collapse and grow to the atomic-cooling limit (ACL) where cooling via atomic hydrogen can induce rapid isothermal collapse at $T\sim10^4$K, potentially building a supermassive star (SMS) of mass $M {\sim} 10^3{-}10^5 {\rm M_\odot}$ \citep{Loeb_1994,Oh_2002, Bromm_2003, Lodato_2006, Shang_2010, Regan_2020}. If these SMSs formed, they would promptly leave behind 'heavy' black hole seeds. Heavy seeds could also be produced by hyper-Eddington accretion onto a Pop III star black hole remnant \citep{Ryu_2016, Inayoshi_2016} and runaway collisions in dense proto-clusters \citep{Boekholt_2018, Tagawa_2020,Escala_2021, Vergara_2022, Schleicher_2022, Pacucci_2025}.

Recent work has focused on developing observational diagnostics. An abnormally high ratio of the luminosity emitted in the He ${\rm II} \ \lambda 1640$ {\it vs.} the H$\alpha$ line \citep{Tumlinson_2000,Oh_2001,Johnson_2009,Johnson_2011} can diagnose the presence of massive black holes in high-redshift sources. Further, unique spectral signatures in spectral lines and broadband colors \citep{Pacucci_2015,Pacucci_2019_DCBH, Nakajima_2022, Inayoshi_2022} and the overmassive black hole relation, where the black hole mass can be comparable or even larger than the total stellar mass, or $M_\bullet/M_* \geq 1$ \citep{Volonteri_2008, Scoggins_2024}, can distinguish heavy {\it vs.} light seeds. This overmassive ratio is several orders of magnitude larger than local black hole to stellar mass relations $M_\bullet/M_* {\sim} 10^{-3}$ \citep{Reines_2015}. The heavy-seed pathway may have little or no star formation prior to the formation of the supermassive star (SMS) and subsequent black hole, so this mass ratio could initially, at the birth of the black hole, be as high as $10^3$. Previous work has focused on estimating the lifetime of this overmassive phase \citep{Scoggins_2022, Scoggins_2024}, which is expected to decay as the host halo forms stars and/or merges with halos that have a ratio approaching the low-$z$ values. However, this relation may continue to exist in the case of halos that experience very few or no mergers, such as in isolated dwarf galaxies, providing a unique opportunity to search for heavy seed descendants \citep{Wassenhove_2010,Greene_2012,Ricarte_2018, Mezcua_2023,Regan_2023, Tremmel_2024}. We note that the overmassive feature at low redshift may sometimes arise from complex environmental interactions, rather than heavy seeds \citep{Weller_2023}.

A local dwarf galaxy, Leo~I, has recently come into focus for this reason. It has been suggested that Leo~I could host an unusually massive black hole \citep{ Rosell_2021, Pacucci_2022, Pascale_2024, Pascale_2025}, with recent work finding a black hole's existence at 95\% confidence with mass $M_\bullet = 3.3 \pm 2.2 {\times} 10^6 $~M$_\odot$ \citep{Bustamante_2021}. With the total stellar mass estimated to be $M_* =5.5{\times} 10^6~{\rm M}_\odot$ \citep{McConnachie_2012}, this black hole appears to be overmassive, with comparable black hole and stellar masses.  This is similar to what is expected for heavy-seed black holes near the time of their formation, and motivates a question: \textit{could Leo~I's be the descendant of a heavy seed formation site?}   In this paper, we address this question, but we caution that \citet{Pascale_2024, Pascale_2025} argue that the black hole mass is lower, few${\times} 10^5$M$_\odot$, and suggest that Leo~I's high DM density can mimic a black hole, and there may be no black hole after all.

Recently, extremely overmassive black holes have also been found at cosmic noon, extending the track of such objects from the epochs probed by JWST (i.e., $4\lsim~z \lsim~10$) towards the local Universe. For example, \citet{Mezcua_2024} found 12 SMBHs that are hosted by low-mass galaxies with masses 1-2 orders of magnitude below the local scaling relations.

Whether Leo~I is truly overmassive or even hosts a black hole is still not settled. Although the low binding energy of dwarf galaxies makes them vulnerable to tidal stripping, Leo~I appears to have had few mergers in its history. It went through a period of intense star formation but has since run out of gas, and star formation has been mostly quenched \citep{Gallart_1999, Pacucci_2023b}. This means that Leo~I has had very little change in black hole and stellar mass over Gyr timescales, making it a promising "fossil" to study the overmassive relation established in the early universe.

To determine the likelihood of Leo~I hosting a heavy seed remnant, we use Monte-Carlo dark matter merger trees and a semi-analytic model previously developed in \citet{Scoggins_2024}. We estimate the chance that a heavy seed could survive mergers and end up in a halo similar to Leo~I, which we dub 'heavy-seed survivor' (HSS). This model estimates the frequency of HSSs in a dwarf galaxy similar to Leo~I by searching halos near $z{\sim} 30$ that avoid early fragmentation and reach the atomic-cooling limit without prior star formation, where runaway collapse forms a SMS and a prompt  heavy-seed remnant black hole. Applying this model to 1,000 merger trees and tracking halos near redshift $z=0$ that share the properties similar to Leo~I allows us to estimate the fraction of Leo-like satellite halos that could have evolved from a heavy seed hosting galaxy (HSS).

The rest of this paper is organized as follows. In \S~\ref{sec:methods} we describe our methods. In \S~\ref{sec:results} we present our results. In \S\ref{sec:discussion} we discuss these findings. Finally, we summarize our conclusions and the implications of this work in~\S~\ref{sec:conclusion}.

\section{Methods}
\label{sec:methods}
In this section, we summarize the generation of our Monte Carlo dark matter halo merger trees, our heavy seed selection criteria within these merger trees, our selection for Leo-like candidates within these heavy seed hosting branches, and our prescription for stellar mass and black hole growth. This work assumes the following cosmological parameters: $\Omega_{\Lambda}= 0.693$, $\Omega_m =0.307$, $\Omega_b = 0.0486$, $\sigma_8 = 0.81$, and $h=0.67$ \citep{Planck_2018}.

\subsection{Monte-Carlo merger trees for dark matter halos}
We generate 1,000 Monte-Carlo merger trees for dark matter halos based on the Extended Press-Schechter theory \citep{Press_1974}, following the algorithm in \citet{Parkinson_2007}, which is a modified version of the algorithm used in the \texttt{GALFORM} semi-analytic galaxy formation model \citep{Cole_2000}. In order to find Leo-like candidates, we look for satellite halos near redshift $z{=}0$ in the branches of a parent halo at redshift $z{=}0$ with a mass and age similar to the Milky Way. This sets the parent mass of our merger tree to $9{\times}10^{11}~{\rm M_\odot}$, at redshift $z{=}0$. We set a redshift step size of $dz{=}0.166$, with a minimum mass and mass resolution of $10^5~{\rm M_\odot}$, where star formation is unlikely in (mini)halos less massive than this \citep{Kulkarni_2021,Schauer_2021}.

\subsection{Identifying heavy-seed sites and Leo candidacy}
Here, we briefly summarize the model of \citet{Scoggins_2024} that was used in this work. To achieve the intermediary SMS and subsequent heavy seed, ${\rm H}_2$ cooling must be either suppressed or offset to prevent fragmentation and star formation before reaching the atomic-cooling limit. This can be achieved through combinations of intense Lyman-Werner radiation (with specific intensity $J_{\rm LW}$) which disassociates ${\rm H}_2$ \citep{Haiman_1997, Dijkstra_2008, Dijkstra_2014, Wolcott_2017}, and dynamical heating via halo mergers (at a rate $\Gamma_{\rm dyn}$) and large baryonic streaming motions ($v_{\rm stream}$) which can prevent gas infall and contraction. If these processes can prevent fragmentation until the atomic-cooling halo stage with $T_{\rm vir}{\sim} 10^4{\rm K}$, the emission of atomic hydrogen will rapidly cool the gas in the halo, leading to isothermal collapse, possibly producing a massive BH seed via a SMS.

In order to estimate the influence of these effects, for every snapshot in the merger tree, we calculate the cooling time $t_{\rm cool}$ and compare it to the Hubble time $t_{\rm Hubble}$. Our model for the cooling time is dependent on Lyman-Werner radiation and dynamical heating. Here, we briefly summarize the calculation of our halos' Lyman-Werner radiation background and dynamical heating effects, and refer the reader to \S~2.2 and \S~2.3 in \citet{Scoggins_2024} for full details. Following equation (5) of \citet{Scoggins_2024}, we calculate the mean Lyman-Werner intensity expected to be experienced by every halo in our merger trees, $\overline{J}_{\rm LW}(M_{\rm halo}, z)$. This captures the mean, but it is expected that the halos that form heavy seeds will experience $J_{\rm LW}$ at the extreme end of the distribution ($J_{\rm LW} {\gsim} 10^3 J_{21}$, e.g. \citealt{Shang_2010, Glover_2015,Agarwal_2016, Wolcott_2017}). To account for the different intensity experienced by every halo, we draw from a numerically determined $J_{\rm LW}$ distribution shown in Fig. 9 of \citet{Lupi_2021}, centered on $\overline{J}_{\rm LW}(M_{\rm halo}, z)$. We do this for every halo above the atomic-cooling threshold (ACT). For halos just above the ACT, we calculate the ratio $\alpha = J_{\rm LW}/\overline{J}_{\rm LW}$, and the progenitors at and below the ACT are estimated to experience $J_{\rm LW} = \alpha \overline{J}_{\rm LW}$. This accounts for the fact that a halo experiencing unusually high (low) LW flux exists in an overcrowded (underdense) region, and presumably the progenitors of this halo experience a similarly higher (lower) $J_{\rm LW}$ flux. We calculate the dynamical heating rate following equation (1) of \citet{Wise_2019}, 
\begin{align}
\label{eqn:heating}
    \Gamma_{\rm dyn} = \frac{T_{\rm halo}}{M_{\rm halo}} \frac{k_{\rm B}}{\gamma -1} \frac{dM_{\rm halo}}{dt},
\end{align} 
for adiabatic index $\gamma = 5/3$. Using $J_{\rm LW}$, among other details listed in \citet{Scoggins_2024}, to calculate the cooling rate, and Eq.~\ref{eqn:heating} to calculate the heating rate, we can derive an estimate for the net effective cooling time in each halo. We note that this model will underestimate the HSS occurrence rate, as it is still possible for SMSs to form in halos where cooling via ${\rm H_2}$ does occur, as long as they experience rapid mass inflows that allow for the formation of a SMS \citep{Hosokawa_2013, Haemmerle_2018, Wise_2019}.

With an estimate for the cooling time for every snapshot, we define two parameters that we explore in this work: The minimum dimensionless cooling-time, $\tau_{\rm cool}$, so that the gas inside halos must have $t_{\rm cool}/t_{\rm Hubble}\geq\tau_{\rm cool}$ to be considered pristine, and the virial temperature that defines the onset of atomic cooling, or the atomic-cooling threshold, $T_{\rm act}$. We consider a halo to be a direct-collapse black hole (DCBH) host candidate at the first snapshot where $T_{\rm vir} \geq T_{\rm act}$ if every progenitor across every branch of this atomic-cooling halo satisfy $t_{\rm cool}/t_{\rm Hubble} > \tau_{\rm cool}$. After reaching $T_{\rm act}$, it is assumed that runaway atomic cooling condenses the cloud and forms an SMS. For a given $\tau_{\rm cool}$ and $T_{\rm act}$, we search for halos that could potentially lead to a heavy seed, and then follow the subsequent evolution of these halos. We filter these halos for Leo-like candidacy by analyzing the DCBH descendants of these branches. To summarize, our process of searching for HSSs follows:

\begin{enumerate}
    \item We vary the temperature of the atomic-cooling threshold, exploring a range of $T_{\rm act}  \ \in \ [0.4,1]{\times} 10^4 K$.
    \item We vary the minimum allowed ratio between the cooling time and the Hubble time for all progenitors before the ACT, $\tau_{\rm cool} = t_{\rm cool}/t_{\rm Hubble}$, exploring $\tau_{\rm cool} \in [0.05, 1]$
    \item For a given $\tau_{\rm cool}$ and $T_{\rm act}$, we look for DCBH descendants in the merger tree within the virial mass range of Leo~I, $(7 \pm 1){\times}\ 10^8 {\rm M_\odot}$ \citep{Mateo_2008}, that exists within the free-fall time of $z=0$. This results in a window of ${\sim}1$ Gyr before $z=0$, or $t_{\rm ff} = \frac{\pi}{2}\frac{d_{\rm Leo}^{1.5}}{\sqrt{2G M_{\rm MW} M_{\rm Leo}}}=1.090$ Gyr for a Leo-MW separation of $d_{\rm Leo} = 250$ kpc.
    \item For the DCBH-host halos with a mass similar to Leo~I within a free-fall time of redshift $z{=}0$, we filter these halos according to their merger history. We require that the Leo-like halo has experienced one to three major mergers between $z=0.1$ and the redshift of the atomic-cooling crossing, as shown by \citet{Ruiz-Lara_2021}, based on the analysis of the star formation history and the age-metallicity relation. This introduces another parameter in our search for HSSs, the minimum mass ratio for a major merger, $q = M_{\rm final}/M_{\rm initial}$. For completeness, we explore the full allowed range of $1\leq q\leq 2$.
\end{enumerate}

Although this version of \texttt{GALFORM} does not explicitly track satellite halos, the Extended Press-Schechter formalism used in this work considers two halos as "merged" when they become closely gravitationally bound. \citet{Tanaka_2009} have highlighted that for a host halo that is more than 20 times more massive than its satellite, the infall time is so long that the satellite is considered "stuck" and likely never merges. This justifies our choice of searching for Leo-like halos within a free-fall time: even though they formally 'merge' in our merger trees at $z=0$, they are likely to remain long-lived satellites.

We consider a halo to be a Leo-like heavy-seed survivor, or HSS, if it meets all of the above criteria. These parameters are summarized in Table~\ref{table_params}.

\subsection{Stellar mass and black hole growth}
We assign stellar masses to our halos following \citet{Wise_2014}, who derived stellar mass versus halo mass statistics from a cosmological hydrodynamical simulation. In their Table~1, they provide log$(M_{\rm vir})$ and log$(M_*)$ statistics for $6.5 \leq\log(M_{\rm vir}/{\rm M}_\odot) \leq 8.5$ in 0.5 dex bins. We interpolate across log$(M_{\rm vir})$ to derive log$(M_*)$ for a given halo mass and apply this to halos with $10^{6.5} \leq M_{\rm halo}/{\rm M_\odot} \leq 10^{8.5}$. We note that these statistics are generated from a simulation that ran until $z=7.3$, but we apply them to halos with redshift $z \geq 6$.

Black holes are assumed to form shortly after the halos reach the ACT. The initial seed black hole masses in\texttt{Renaissance}, a suite of cosmological radiation-hydrodynamic and N-body simulations \citep{OShea_2015, Xu_2016}, are estimated to fall within the range $10^4  {\rm M_\odot}\leq M_\bullet \leq 10^6 {\rm M_\odot}$ where the gravitational collapse to a SMBH is triggered by a relativistic instability. We note that a re-simulation of two of the atomic-cooling halos in the \texttt{Renaissance} suite found lower SMS masses of $M{\approx}10^2-10^{4} {\rm M_\odot}$ \citep{Regan_2020b}, with higher $J_{\rm LW}$ yielding a higher mass.  However, the halos in this re-simulation experienced much smaller $J_{\rm LW}$ fluxes (${\sim}10 J_{21}$) than we investigate in this work $({\sim} 10^3 J_{21})$, so we expect our seeds to be more massive. We estimate the initial black hole mass to be some fraction of the baryon mass, $M_0 = f_{\rm cap}\left(\frac{\Omega_b}{\Omega_m}\right)M_{\rm halo}$, with this fractional cap set to $f_{\rm cap} = 0.05$. This typically yields black holes with masses ${\sim} 10^4 {\rm M_\odot}$. The growth of these black holes is assumed to follow the Eddington rate
\begin{align}
   &\dot{M}_{\rm bh} = \frac{L_{\rm edd} }{\epsilon c^2} = \frac{4\pi G \mu m_{\rm p} M_\bullet}{\sigma_{\rm T}c \epsilon} = \frac{M_\bullet}{\tau_{\rm fold}},
\end{align}
with speed of light $c$, gravitational constant $G$, mean molecular weight $\mu$ ($\mu {\sim} 0.6$ for fully ionized primordial H + He gas), proton mass $m_{\rm p}$, Thomson cross section $\sigma_{\rm T}$ and radiative efficiency $\epsilon$. This leads to a black hole mass given by $M_\bullet(t) = M_0\exp(t/\tau_{\rm fold})$ with e-folding time $\tau_{\rm fold} =  (\sigma_{\rm T} c \epsilon)/(4\pi \mu G m_{\rm p}) \approx  $ 450$\epsilon$ Myr. Assuming a radiative efficiency $\epsilon \approx 0.1$, we set $\tau_{\rm fold} =45$ Myr. We additionally quench black hole growth when the mass of the black hole exceeds the prescribed fraction of the baryonic mass in the halo, capping $M_\bullet \leq f_{\rm cap}\left(\Omega_{\rm b}/\Omega_{\rm m}\right)M_{\rm halo}$. To summarize, our simple model governs black hole formation and growth through $f_{\rm cap}, \tau_{\rm fold}$, $M_{\rm halo}$, and $M_0$ (which is determined by $f_{\rm cap}$ and $M_{\rm halo}$). These parameters are summarized in Table~\ref{table_params}. We start the growth of our black holes immediately after their formation.

\begin{table}
\centering
\begin{tabular}{|| c | c | p{3cm} ||} 
 \hline  
\hline
Parameter & Value/Range & Description\\
\hline
$T_{\rm act}$ & $[4000, 10^4]$K & The virial temperature that defines the onset of atomic cooling.  \\
\hline
$\tau_{\rm cool}$ & 
$[0.05,1]$ & The minimum cooling time $t_{\rm cool}/t_{\rm Hubble}$ required for a halo to be considered pristine.  \\

\hline
$q$ & $[1.0,2.0]$  & The minimum mass ratio $M_{\rm final}/M_{\rm initial}$ that defines a major merger. \\
\hline
$f_{\rm cap}$ & 0.05 & The maximum fraction of baryonic material available for black hole formation.  \\
\hline
$\tau_{\rm fold}$ & 45 Myr & Black hole e-folding time.  \\

\hline
$M_0$ & $f_{\rm cap}\left(\frac{\Omega_b}{\Omega_m}\right)M_{\rm halo}$ & 
The initial black hole mass. This equation also governs the maximum black hole mass during evolution.\\
\hline
\hline
\end{tabular}
\caption{The values or ranges of several parameters used in this work. These parameters control black hole formation and growth and the HSS criteria.}
\label{table_params}
\end{table}

\section{Results}
\label{sec:results}

\begin{figure*}
\includegraphics[width=0.95\columnwidth]{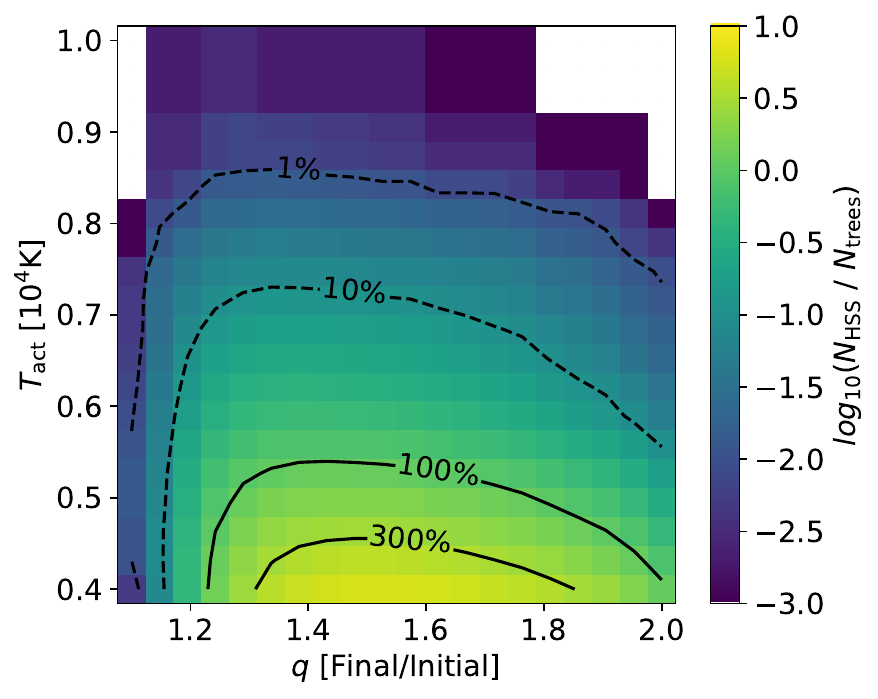}
\includegraphics[width=0.95\columnwidth]{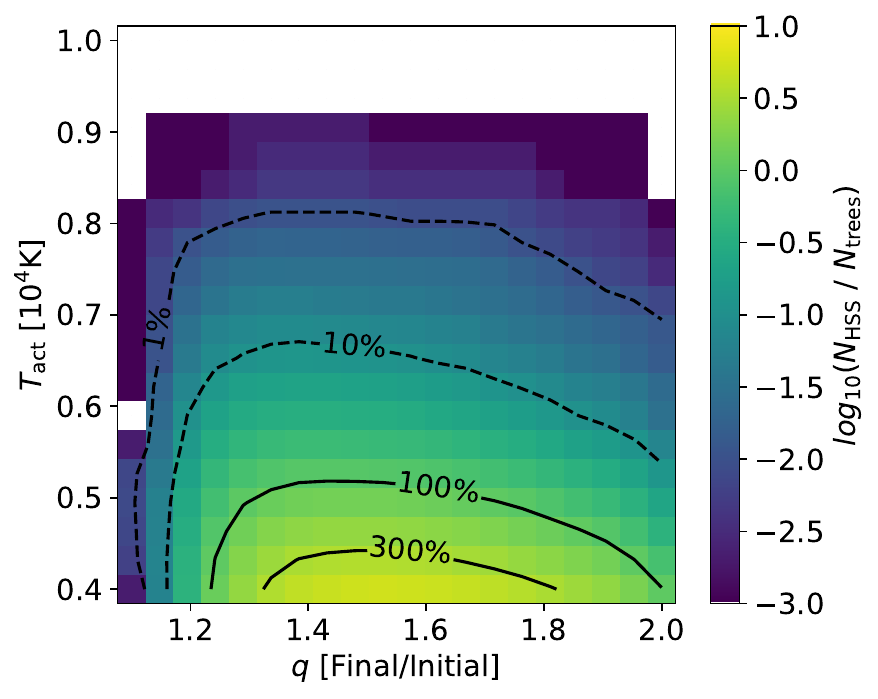}

\includegraphics[width=0.95\columnwidth]{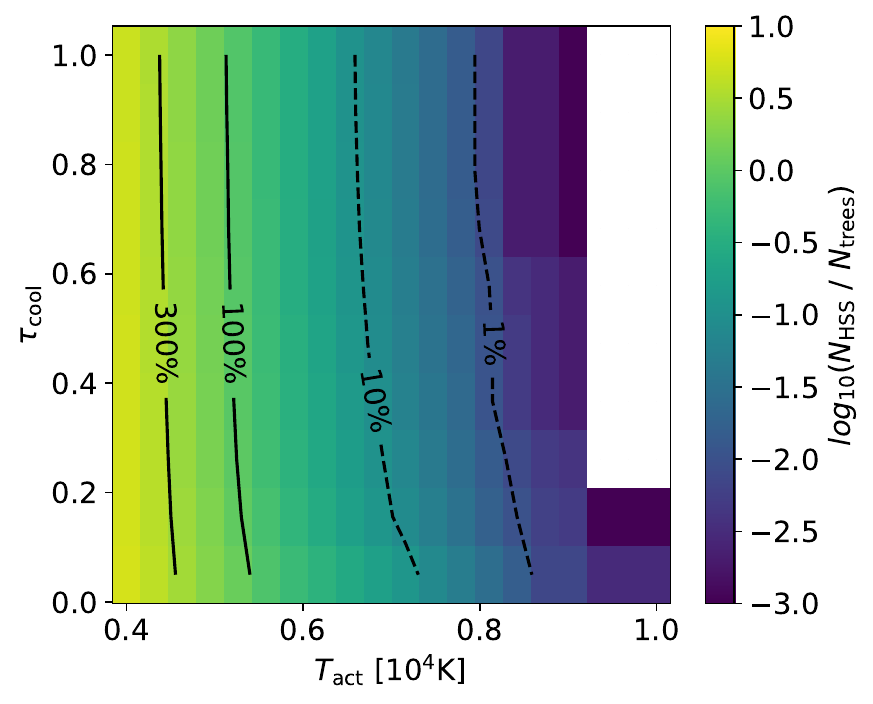}
\includegraphics[width=0.95\columnwidth]{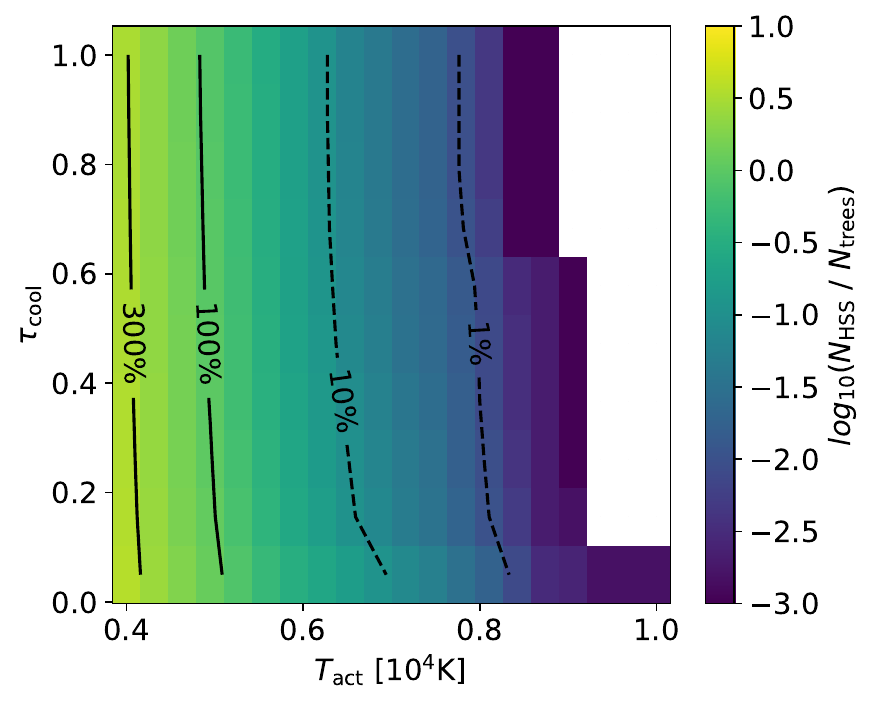}

\includegraphics[width=0.95\columnwidth]{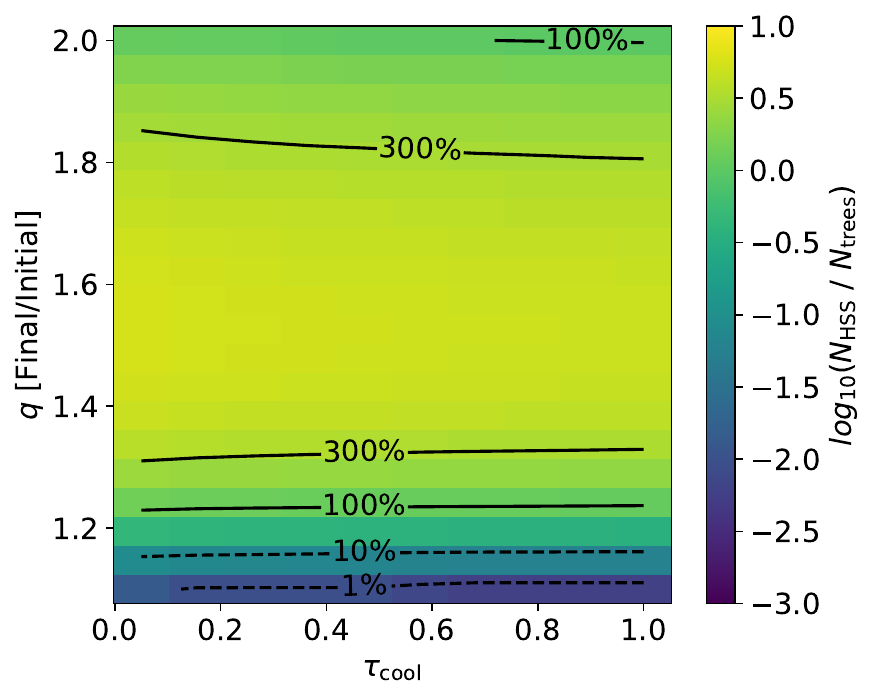}
\includegraphics[width=0.95\columnwidth]{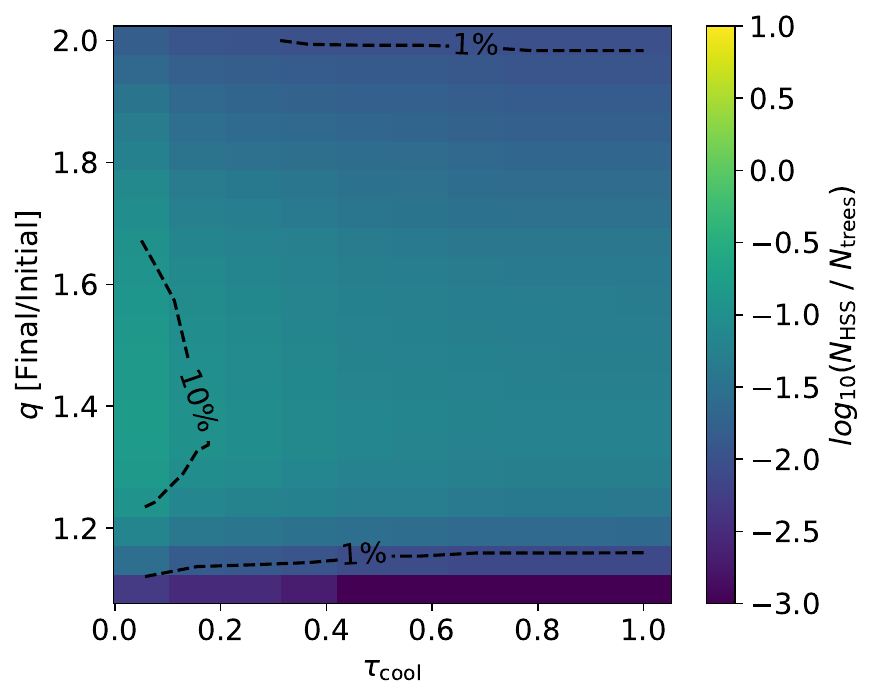}
\caption{We show $\log_{10}(N_{\rm HSS}/N_{\rm trees})$, the frequency of occurence of heavy-seed survivors (HSSs), or a Leo-like satellite halo hosting a DCBH descendant, as a function of the criteria that determine HSS candidacy.  This criterion includes the minimum halo mass ratio that determines what constitutes a 'major' merger, $q$, the virial temperature for the onset of atomic cooling, $T_{\rm act}$, and the minimum allowed cooling time ratio to avoid star-formation, $\tau_{\rm cool} = t_{\rm cool}/t_{\rm Hubble}$. An HSS is formed if a Leo-like halo hosts a DCBH (requiring the progenitors of a subhalo at $T_{\rm vir}{=}T_{\rm act}$ to avoid any prior star formation, 
and that after forming the DCBH, the subhalo experiences one to three 'major' mergers with mass ratio above $q$. Each panel explores two parameters at a time, with the left column showing the value of $\log_{10}(N_{\rm HSS}/N_{\rm trees})$ selected to be the maximum for any value of the third parameter, and the right column showing the median across this third parameter. Lines denote $1\%$, $10\%$, $100\%$, and $300\%$ rates for HSS frequency in our trees. White spaces indicate that no HSS was found among our $N_{\rm trees} = 1,000$ merger trees, so the probability of an HSS existing for a given parameter combination is less than $10^{-3}$. In some cases, there can be three or more HSSs per tree. The HSS frequency has a soft dependence on $\tau_{\rm cool}$, as our halos experience large values of $J_{\rm LW}$ with $\tau_{\rm cool}$ typically greater than 1, so increasing $\tau_{\rm cool}$ does not significantly decrease the number of HSSs. There is a non-linear dependence on $q$, where a large value of $q$ can result in too few major mergers, violating the Leo-like criteria, and a small value of $q$ results in too many mergers. The results have the strongest dependence on $T_{\rm act}$, as decreasing this value increases HSS rates significantly.}
\label{fig:prob_plots}
\end{figure*}

In Fig. \ref{fig:prob_plots}, we show $\log_{10}(N_{\rm HSS}/N_{\rm trees})$, the HSS frequency of occurence for our $N_{\rm trees}=1000$ dark matter merger trees. We show the results as a function of the three parameters explored here, $\tau_{\rm cool}$, $T_{\rm act}$, and $q$. We include six plots, each showing the expected number of HSS candidates per tree across two of the three parameters. For a given value along the x and y axes, the left column shows the maximum number of HSSs across the third parameter not shown on the axes. The right column shows the median number of HSSs along this third parameter. We mark the 300\%, 100\%, 10\%, and 1\% lines for the expected number of HSSs per tree, shown in black. White space indicates that no HSS candidate was found in our $N_{\rm trees} = 1,000$ dark matter merger trees, meanign that the probability of an HSS for that combination of parameters is ${<}10^{-3}$.

The top row shows the expected number of HSSs across the minimum mass ratio for mergers, $q$, and the threshold halo virial temperature $T_{\rm act}$ for atomic-cooling, with the left panel showing the maximum value for each pair across $\tau_{\rm cool}$ and the right panel showing the median. We find that, in general, decreasing $T_{\rm act}$ increases the frequency of HSSs. This is because a smaller value of $T_{\rm act}$ focuses on an earlier, lower mass stage in the halo's evolution, reducing the chances for cooling to occur and resulting in fragmentation. However, the frequency is not monotonic across the minimum halo merger mass ratio, and we find that ratios between $q=1.3$ and $1.6$ result in the highest HSS occurrence rates. Although a smaller value of $q$ is less strict, this tends to result in more mergers than our maximum allowed number of mergers, 3 (recall that we require between 1 and 3 major mergers). For the case of $T_{\rm act} {\sim} 0.4{\times} 10^4$K and $q {\sim} 1.33$, nearly all of our parent halos host one or more HSS. As $T_{\rm act}$ approaches $ 0.7{\times} 10^4$K, the HSS frequency is reduced to $\sim$10\%, or roughly 100 HSSs among our 1,000 merger trees. For a larger value, $T_{\rm act}{\sim} 0.9 {\times} 10^4$K, HSS frequency is less than 1\%. The results appear to be similar when comparing the maximum value (left) to the median value (right) across the third parameter, $\tau_{\rm cool}$, signaling that this parameter plays a less significant role in the frequency of HSSs.

The middle rows show the results as a function of $T_{\rm act}$ and $\tau_{\rm cool}$. As expected by the nature of picking the maximum value of $N_{ \rm HSS}$ across the third parameter, $q$, the dependence on $T_{\rm act}$ is similar to the top row, where the probability increases with decreasing $T_{\rm act}$. Although a lower $\tau_{\rm cool}$ for a fixed $T_{\rm act}$ slightly increases the frequency of HSSs, this highlights the weak dependence on $\tau_{\rm cool}$. This is likely due to the large Lyman-Werner radiation experienced by the heavy-seed sites (see Fig. 3 of \citealt{Scoggins_2024}),. These sites experience abnormally large values of $J_{\rm LW}$, typically being large enough to cause the cooling time to exceed the Hubble time for the majority of our halos. This means that a lower value of $\tau_{\rm cool}$ yields little increase in HSS frequency.

The bottom row explores the HSS frequency as a function of $\tau_{\rm cool}$ and $q$. On the left, showing the maximum value across $T_{\rm act}$, the majority of the parameter space finds most halos that host an HSS, although this maximum value is likely always achieved for a low value of $T_{\rm act}$. Again, there is little dependence on $\tau_{\rm cool}$, the results across $q$ peak near ${\sim} 1.5$. A value of $q$ below $1.2$ results in so many 'major' mergers that we regularly exceed our maximum allowed number of three, resulting in almost no HSSs. The right panel, which shows the median HSS number across $T_{\rm act}$, reveals an interesting dependence on $\tau_{\rm cool}$ and $q$, where in this space most of the peak HSS frequency is near $q{\sim} 1.4$ for an extremely small $\tau_{\rm cool}$. For larger values of $\tau_{\rm cool}$, for the optimal $q$, the HSS frequency drops to ${\sim} 5\%$.

\begin{table}
\centering
\begin{tabular}{||c | c | c | c ||} 
 \hline  
\hline
Strictness & $q$ & $T_{\rm act}$ [K] & $\tau_{\rm cool}$\\
\hline
Least & 1.4 & 5,000 & 0.5 \\
\hline
Medium & 1.4 & 7,000 & 0.75 \\

\hline
Most & 1.4 & 9,000 & 1.0 \\
\hline
\hline
\end{tabular}
\caption{The three cases shown in Fig.~\ref{fig:num_cands} that define the minimum criteria required for HSS candidacy.}
\label{table1}
\end{table}

\begin{figure}
\includegraphics[width=0.95\columnwidth]{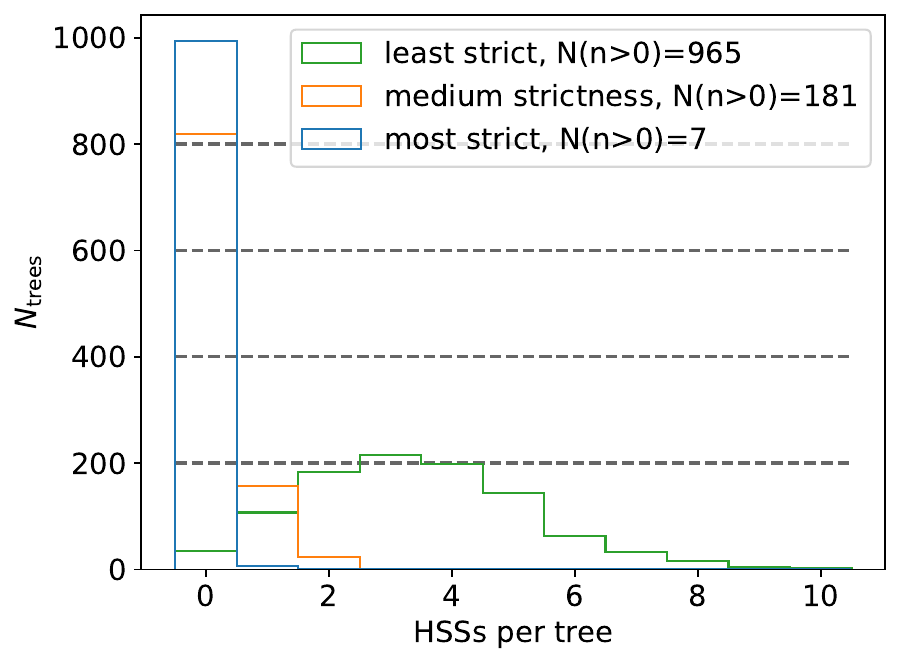}

\caption{The distribution of the number of HSSs in our 1,000 merger trees. The three cases represent varying degrees of strictness in our three parameters, $q$, $T_{\rm act}$, and $\tau_{\rm cool}$, with the values of these parameters listed in Table~\ref{table1}. In the least strict case (green), $96.5\%$ of our trees hold at least one HSS, though it is much more common for the trees to have several HSSs. One tree holds as many as 10 HSSs. For the medium case (orange), $18.1\%$ of our trees hold an HSS, with two being the upper limit of HSS per tree. In the strictest case (blue), only $0.7\%$ of our trees have HSSs. This stricter case is the most physically realistic, though future work will be needed to put tighter constraints on the most appropriate $T_{\rm act}$ for rapid gas collapse in atomic-cooling halos.}
\label{fig:num_cands}
\end{figure}

Our results for the total number of Leo-like satellite halos that host a heavy seed, or heavy-seed survivors (HSS), for three varying levels of strictness are shown in Fig.~\ref{fig:num_cands}. The criteria that define these three cases are summarized in Table~\ref{table1}. As the number of HSSs depends nonlinearly on $q$, we have fixed it to the approximately optimal value of $q=1.4$, allowing our cases to explore the dependence on the other two parameters, $\tau_{\rm cool}$ and $T_{\rm vir}$. For the least strict scenario (green), the majority of our trees (965 out of 1,000) end up with several HSS candidates, up to 10 HSSs in a single tree. For the medium-strict case, 181 out of 1,000 of our trees harbor an HSS. Of these, most have only a single HSS, with a small fraction hosting up to two but no more. Finally, in the strictest case, very few HSSs remain. Only 7 of our 1,000 trees have a (single) HSS. Overall, these numbers suggest that it is not unreasonable for Leo~I to host a heavy-seed black hole descendant, although to make this less uncommon, our criteria need to be on the less strict side. However, even in the strictest case we imposed, Leo~I could still host a heavy seed.

\begin{figure}
\includegraphics[width=0.95\columnwidth]{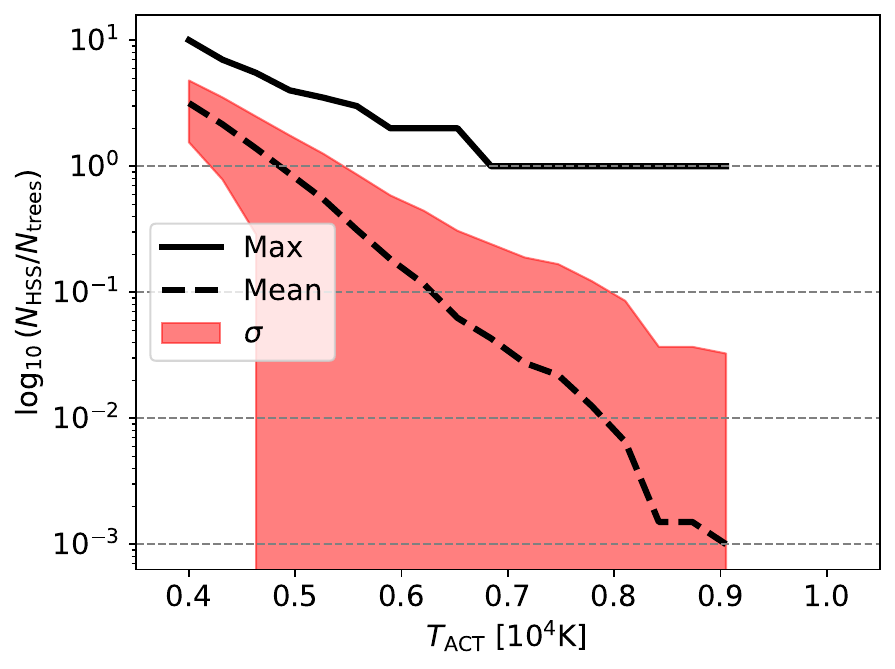}
\caption{We show the HSS frequency as a function of $T_{\rm act}$, the most influential parameter for HSS candidacy. For every tree, we take the median value (dashed) across $q$ and $\tau_{\rm cool}$ for each target $T_{\rm act}$, then average these results across the 1,000 merger trees. We also show the maximum number of HSSs (solid) and the standard deviation (red) across the 1,000 trees. For low values of $T_{\rm act}$, trees typically host three HSS, up to a maximum of 10. This frequency declines rapidly with increasing $T_{\rm act}$, and there are no HSSs for $T_{\rm act} > 9,000$ Kelvin. The plateau in the maximum number of HSSs signals that from $T_{\rm act} {\sim} 7,000$K up to $T_{\rm act} {\sim} 9,000$K, there is no more than one HSS in the median value across $q$ and $\tau_{\rm cool}$.}
\label{fig:n_vs_t}
\end{figure}

In Fig.~\ref{fig:n_vs_t} we compare the number of HSSs against $T_{\rm act}$, the most influential parameter for determining how many HSSs end up in a MW-like halo. For each merger tree and target $T_{\rm act}$, we select the median number of HSSs across $q$ and $\tau_{\rm cool}$. Then, with the median value for each tree, we show the mean (dashed), maximum (solid), and r.m.s. ($\sigma$; red) across the 1,000 merger trees. This can be thought of as a summary of the information in the top four panels of Fig. \ref{fig:prob_plots}. For low values of $T_{\rm act}$, we find that there are typically three or more HSSs per tree, but as we approach a more realistic value $T_{\rm act}{\sim} 7,000$K, there are roughly 50 HSS across the 1,000 trees. Above $9,000$K, there are no HSSs. The solid black line shows that there are a maximum of 10 HSSs in one tree across $q$ and $\tau_{\rm cool}$, though it plateaus to one HSS near $T_{\rm act}{=} 7,000$K and zero above $T_{\rm act}{\sim} 9,000$K.

\begin{figure*}
\includegraphics[width=0.9\columnwidth]{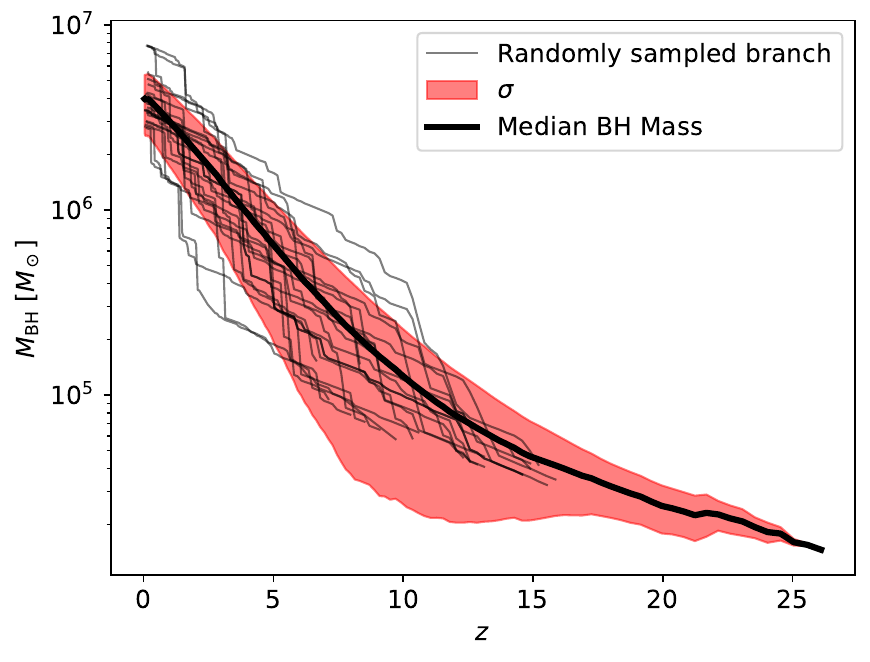}
\includegraphics[width=0.9\columnwidth]{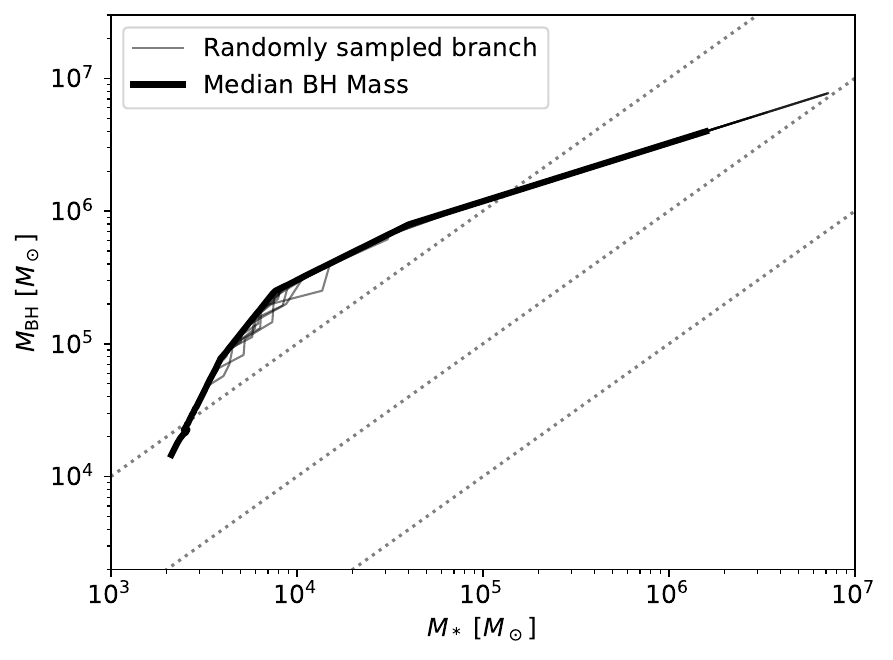}

\caption{{\it Left:} The evolution of the masses of black holes born as heavy seeds in our models. We show the median black hole mass (black), a few randomly sampled branches of the merger tree of a Milky-Way-like halo (gray) and the black hole mass standard deviation {\it vs.} redshift (red). The median black hole mass near redshift $z{=}0$ is $M_\bullet{\sim}4{\times} 10^6 {\rm M}_\odot$, comparable to the estimate by \citet{Bustamante_2021}. The earliest heavy-seed survivors (HSSs) at high redshift have an initial mass of $M_\bullet{\sim} 10^4~{\rm M}_\odot$. {\it Right:} The black hole {\it vs.} stellar mass for our HSSs. The HSSs are initially overmassive, with $M_\bullet/M_* {\sim} 10$, and terminate with $M_\bullet/M_* {\sim} 1$. }
\label{fig:bh_results}
\end{figure*}

In Fig. \ref{fig:bh_results}, we show the evolution of our HSS candidate black holes for branches that resulted in HSS candidacy for the least strict case. Although such a low $T_{\rm act}$ for atomic cooling and DCBH formation is unlikely to be realistic, as the onset of atomic cooling typically requires $T_{\rm act} \gtrsim 8,000$, the halo evolution is similar in the stricter scenarios, and this allows us to see a larger sample size of black holes. In the left panel, we show black hole mass {\it vs.} redshift, with the median black hole mass shown in black, the standard deviation shown in red, and randomly sampled branches shown in gray. The earliest black hole formation time is near redshift $z=25$, with an initial heavy-seed mass of ${\sim}10^4~{\rm M}_\odot$. The black holes grow to have a median mass of ${\sim}4{\times} 10^6~{\rm M}_\odot$ near redshift $z{=}0$, which closely matches the mass of Leo~I's black hole derived by \citet{Bustamante_2021}. In the right panel, we show the evolution of our black hole mass against the stellar mass of their host halos. As our black holes grow, they are limited to a fraction of the gas reservoir available in their halo, or $5\%$ of the total baryonic matter. These grow quickly enough that most will soon reach this cap, meaning that black hole mass and stellar mass (which we estimate according to the halo mass) are tightly correlated in this model. Near redshift $z=0$, most of these satellites have a final BH-to-stellar mass ratio of $M_\bullet/M_*{\sim} 1$, similar to the ratio observed in Leo~I \citep{Pacucci_2023b}.

\section{Discussion}
\label{sec:discussion}
This investigation suggests that the presence of a heavy-seed descendant in Leo~I, and in other dwarf satellite galaxies similar to Leo~I, is plausible under moderately relaxed DCBH formation conditions. Although the most restrictive cases only yield ${\sim} 1\%$ of the MW merger trees as HSS hosts, the fraction grows rapidly as the conditions are relaxed, resulting in $18\%$ and $96\%$ HSS frequency for medium and extremely relaxed conditions. This suggests that the main uncertainty in HSS frequency lies in the physics of early halo cooling and fragmentation.

The weak dependence on the dimensionless cooling time, $\tau_{\rm cool}$, suggests that halos meeting the heavy-seed criteria are typically subjected to an intense $J_{\rm LW}$ flux. For the majority of these halos, $t_{\rm cool} \geq t_{\rm Hubble}$ throughout most of their early evolution, meaning that increasing $\tau_{\rm cool}$ from 0.05 to 1.0 has little influence on HSS frequency. Our mass ratio, $q = M_{\rm final}/M_{\rm initial}$, is used to keep the merger history of our candidates similar to Leo~I, so that this criterion does not constrain dwarf galaxies hosting an HSS more generally. The results derived here for a given $\tau_{\rm cool}$ and $T_{\rm act}$ can therefore be thought of as a lower limit for HSS frequency in dwarf galaxies in general. For Leo-like halos, we find that $q{\sim} 1.4$ optimizes the HSS frequency, where a large value results in very few or almost no mergers, and a smaller value can result in so many mergers that the Leo-like criteria is difficult to achieve.

The dominant parameter in this work is the threshold halo virial temperature $T_{\rm act}$ for atomic-cooling. Decreasing $T_{\rm act}$ below $9,000$K dramatically increases the HSS frequency, implying that even small deviations from the canonical $10^4$K threshold can shift the outcome by orders of magnitude. Previous work has suggested that collapse could occur at lower virial temperatures, near $T_{\rm vir}{\sim}8,000$K \citep{Regan_2020c}. At this virial temperature, we derive a HSS frequency of ${\sim} 1\%$ (Fig.~\ref{fig:n_vs_t}). Although our scenarios with $T_{\rm act} \leq 8,000$K are optimistic, they cannot be ruled out. Given the importance of this parameter, we here note several related points.

Our merger trees provide only the mass and redshift of halos, and we assign a $T_{\rm vir}$ following equation (26) of \citet{Barkana_2001}. This value reflects the average depth of the corresponding gravitational potential in the spherical-collapse model, which would roughly correspond to the gas temperature in hydrostatic equilibrium.  However, the true gas temperature in a cosmological halo with given mass $M_{\rm halo}$ at redshift $z$ will vary from halo to halo with a significant scatter, and is not necessarily equal $T_{\rm vir}$ even on average. The relationship between the true gas temperature and the virial temperature, and the scatter of the gas temperature for similar atomic-cooling halos has not yet been fully investigated (see \citealt{Shang_2010} for some related discussion). Further, the virial temperature represents an idealized scenario that does not consider effects that may cause rapid cooling and collapse to occur at lower temperatures. For example, in the \texttt{Renaissance} simulations, \citet{Wise_2019} have identified two halos in which ${\rm H_2}$ cooling did occur and reduced the gas temperature below the atomic-cooling value. Nevertheless, these halos experienced rapid mass inflows towards their core, above the critical value of $\sim0.05~{\rm M_\odot~yr^{-1}}$ required for SMS formation~\citep{Hosokawa_2013,Haemmerle_2018}. In our work, for HSSs not to be exceedingly rare, we need atomic cooling to set in near a virial-temperature threshold of $T_{\rm act} {\sim} 7,000$K, and given the uncertainties discussed above, we can expect that to occur for some unknown fraction of the halos. This highlights the importance of follow-up work that investigates non-idealized gas dynamics for the onset of atomic cooling, considering effects such as turbulence, inflows, and anisotropies as well as clarifying the scatter in $T_{\rm vir}$ for a given mass and redshift and the relationship between $T_{\rm vir}$ and gas temperature.

If Leo~I indeed hosts a $M_\bullet{\sim} 10^6~{\rm M}_\odot$ black hole, its current BH-stellar mass ratio is consistent with our evolved HSS predictions, $M_\bullet/M_* {\sim} 1$. The long quenching timescale \citep{Gallart_1999} and lack of mergers \citep{Pacucci_2023b} likely froze its mass ratio near the initial overmassive state, as we have found the ratio to slightly decrease but at a significantly slower rate than is typical for most DCBHs \citep{Scoggins_2022, Scoggins_2024}. 

We note that after completing the analysis for this work, another local dwarf galaxy, Segue 1, was recently modeled as having an extremely overmassive black hole \citep{Lujan_2025}. The black hole is estimated to have mass $(4.5 \pm 1.5) \times 10^5 {\rm M}_\odot$ and comparably very little stellar mass, $M_* {\sim}10^3~ {\rm M}_\odot$ \citep{Geha_2008}. This results in a BH to stellar mass ratio of $M_\bullet/M_* {\sim} 10^2$ and further supports the possibility that Leo~I and other dwarf galaxies, are excellent testing grounds for the $M_\bullet/M_*$ relations established in the early universe. Future work using either simulations or observational data targeting dwarf systems with similar overmassive ratios at low redshift could directly test this heavy-seed scenario. 

Finally, we highlight a previous work that also searches for massive black hole descendants in satellite galaxies. In \citet{Wassenhove_2010}, they investigate the occupation fraction for both massive and Pop. III remnant black holes in Milky Way satellites, not necessarily those that are similar to Leo~I. They consider the effects of mass inflow on the formation of these massive seeds rather than explicitly modeling cooling and heating. They find that these massive black holes occupy less than a few percent of the dwarf galaxies orbiting the Milky Way, similar to the expected number of HSSs in Leo-like galaxies found in this work.

\section{Conclusions}
\label{sec:conclusion}

Dwarf galaxies are thought to be ideal sites to probe potentially preserved $M_\bullet/M_*$ relations from high-$z$ heavy-seed sites, with the local dwarf galaxy Leo~I being one such exciting heavy-seed progenitor candidate. This work investigated this possibility by using Monte Carlo dark matter merger trees and the model developed in \citet{Scoggins_2024}, to search for 'heavy-seed survivors', or DCBH sites that end up in a satellite halo similar to Leo~I, orbiting a galaxy similar to the Milky Way. We measure the HSS occurrence frequency as a function of the minimum allowed dimensionless cooling time, $\tau_{\rm cool}$, the halo virial temperature that defines the onset of rapid gas collapse in atomic-cooling halos, $T_{\rm act}$, and the minimum halo mass ratio for major mergers, $q$.

We find that it is feasible for Leo~I to be a descendant of a heavy seed formed at $z \geq 20$, although this possibility is strongly dependent on the exact halo virial temperature at which atomic cooling begins, $T_{\rm act}$, and is weakly dependent on $q$ and $\tau_{\rm cool}$. Idealized conditions have set the canonical virial temperature for the onset of atomic cooling to $T_{\rm act}=10^4$K, though recent work has found that this could be lower, closer to $8,000$K \citep{Regan_2020c}. We find a HSS frequency of ${\sim} 1\%$ using this threshold temperature. Future work should aim to refine the details for the onset of rapid gas collapse in atomic cooling halos, determining the scatter of the specific temperature at which atomic cooling begins, $T_{\rm act}$, as well as the discrepancy between the estimated virial temperature and the gas temperature at the center of the halo when it begins to collapse. In reality, the gas will not necessarily be in hydrostatic equilibrium in the halo's potential. Further, even if ${\rm H_2}$ cooling does occur before reaching $T_{\rm act}$, it is still possible to form a SMS, as long as the mass inflow exceeds the critical value of $\sim0.05~{\rm M_\odot~yr^{-1}}$ required for SMS formation~\citep{Hosokawa_2013,Haemmerle_2018, Wise_2019}. Therefore, our estimates for the HSS frequency should be considered conservative.

Overall, our results are consistent with Leo~I hosting a black hole of mass ${\sim} 10^6~{\rm M}_\odot$, with a black hole to stellar mass ratio of $M_\bullet/M_* {\sim} 1$. This suggests that Leo~I, and other similar dwarf galaxies that experienced few or no major mergers in their history, will preserve the overmassive relation and serve as local probes into high-redshift environments.

\section*{Acknowledgements}
ZH acknowledges support from NSF grant AST-2006176. 
FP acknowledges support from the Black Hole Initiative at Harvard University, which is funded by grants from the John Templeton Foundation and the Gordon and Betty Moore Foundation.  
Merger tree generation and analysis were performed with NSF’s ACCESS allocation AST-120046 and AST-140041 on the Stampede3 resource. The freely available plotting library matplotlib \citep{Hunter_2007} was used to construct the plots in this paper.

\section*{Data Availability}
The code used to analyze the merger trees and generate figures for this manuscript is available at this \href{https://github.com/mscoggs/merger_tree_exploration}{github repository}. All other data will be shared on reasonable request to the corresponding author.

\bibliographystyle{mnras}
\bibliography{references}
\bsp



\label{lastpage}
\end{document}